\documentclass[aps,prl,twocolumn,superscriptaddress]{revtex4-1}

\usepackage{hyperref} 
\usepackage{graphicx}
\usepackage{amsmath}
\DeclareGraphicsExtensions{.png,.jpg,.eps}
\usepackage{xcolor}

\begin{document}

\author{J. L. Lado}
\affiliation{Institute for Theoretical Physics, ETH Zurich, 8093 Zurich, Switzerland}
\author{M. Sigrist}
\affiliation{Institute for Theoretical Physics, ETH Zurich, 8093 Zurich, Switzerland}
\title{Two-dimensional
topological superconductivity with
antiferromagnetic insulators}

\begin{abstract}
Two-dimensional topological superconductivity has attracted great interest due
to the emergence of Majorana modes bound to vortices and propagating 
along edges. However, due to its rare appearance in natural
compounds, experimental realizations rely on a delicate artificial
engineering involving materials with helical states, magnetic fields and conventional
superconductors. Here we introduce an alternative path using a class of three-dimensional antiferromagnet 
to engineer a two-dimensional topological superconductor. Our proposal
exploits the appearance of solitonic states at the interface between a
topologically trivial antiferromagnet and a conventional superconductor, 
which
realize a topological
superconducting phase
when their spectrum is gapped by intrinsic spin-orbit coupling.
We show that these interfacial states do not require fine-tuning, but are
protected by asymptotic boundary
conditions. 
\end{abstract}


\maketitle

\begin{figure}[h!]
 \centering
                \includegraphics[width=\columnwidth]{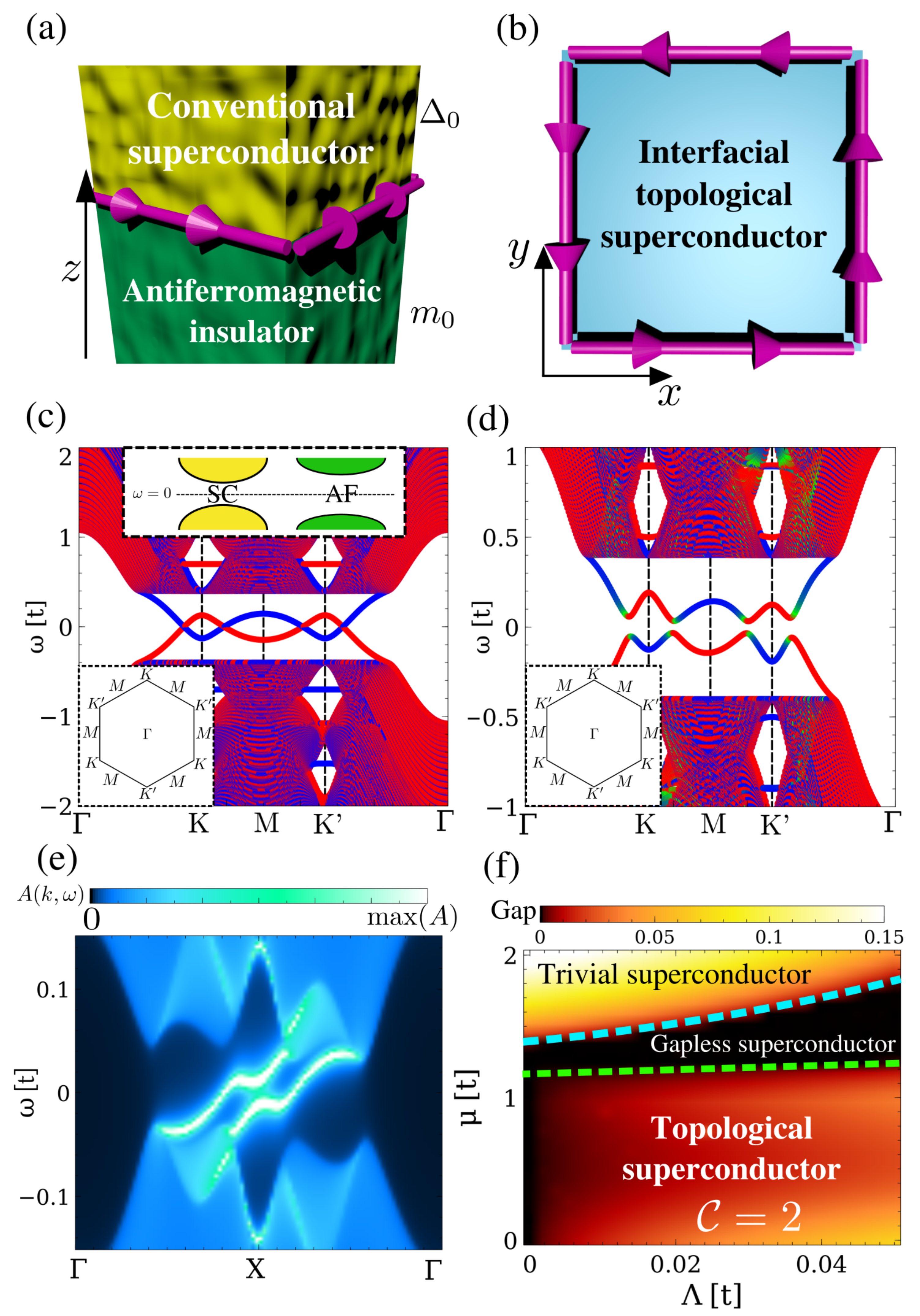}

\caption{ (a) Schematic graph of a three-dimensional
superconducting-antiferromagnet heterostructure,
where a two-dimensional topological superconductor
emerges at the interface (b). Two branches of sub-gap quasiparticle excitations
appear at the interface which have zero-energy modes in the absence of spin-orbit coupling
(c) and are gapped for non-vanishing spin-orbit coupling (d). In the latter case
the system is topological with a Chern number $\mathcal{C}=2$ leading to two
chiral edge modes (e). This topological phase is robust and exists in a wide
parameter range besides a gapless and a trivial superconducting phase (f). 
For the panels (c-d) we consider a sharp interface ($W=0$), with $ t'=t $, $\Delta_0=0.4t$ and $m_0=0.7t$. 
}
\label{fig1}
\end{figure}

Topological matter represents one of the most intriguing frameworks to realize unconventional physics, 
due to $d-1$-dimensional excitations originating from topological properties of the $d$-dimensional systems.\cite{RevModPhys.82.3045,PhysRevB.84.125132,PhysRevB.23.5632} 
This allows to realize electronic spectra in solid state platforms which resemble those found for some elementary
particles in high energy physics and exhibit often an even larger variety. 
Topologically non-trivial band structures
give rise to chiral modes in Chern
insulators,\cite{Liu2016} helical modes
in quantum spin Hall insulators\cite{PhysRevLett.95.226801} 
and Majorana modes in topological superconductors.\cite{leijnse2012introduction,kitaev2001unpaired,PhysRevLett.100.096407}
In particular, Majorana zero-energy modes in one-dimensional (1D)
topological superconductors have fostered intense research efforts both in their
detection\cite{PhysRevX.5.041042,PhysRevX.3.021007,ilan2014detecting} and manipulation,\cite{PhysRevLett.103.107002,li2016manipulating,PhysRevB.85.020502,PhysRevB.85.144501} motivated by their potential for topological quantum computing. \cite{RevModPhys.80.1083,alicea2011non}
Yet, one of the biggest challenges is that nature lacks materials with 1D topological superconductivity, and thus,
the only hope for its experimental realization relies on an artificial engineering in nano-structures
\cite{nadj2014observation,mourik2012signatures,deng2012anomalous,PhysRevLett.110.126406,PhysRevB.84.195442,PhysRevB.88.020407,PhysRevLett.105.077001,PhysRevLett.114.236803}.

Two-dimensional (2D) topological superconductors
share the exciting phenomena of their 1D counterparts,
while providing additional flexibility. On the one hand, Majorana bound states can be found in vortex cores,
\cite{PhysRevLett.86.268,PhysRevLett.116.257003,PhysRevLett.111.136401} 
which display properties of interest for 
topological quantum computing.\cite{PhysRevB.73.220502}
On the other hand, propagating excitations at edges may allow the exploration of the physics of
supersymmetry associated with interacting Majorana fermions.
\cite{PhysRevLett.119.107202,PhysRevLett.115.166401,PhysRevLett.117.166802,PhysRevB.92.235123}
However, natural 2D topological superconductors are
rather elusive,\cite{RevModPhys.75.657}
rendering artificial engineering of 2D topological superconductors an important milestone, very much like their 1D counterparts.
This further motivated extensions of the original mechanisms for 1D topological superconductivity
to two dimensions, based on topological
insulators,\cite{PhysRevLett.100.096407} Yu-Shiba-Rusinov
lattices\cite{PhysRevLett.114.236803}
or 2D electron gases with Rashba spin-orbit coupling.\cite{PhysRevLett.104.040502}

In this letter we introduce an alternative route to create topological
superconductivity, exploiting an interface between two bulk ordered phases.
Our proposal consists of a heterostructure formed by a insulating bulk antiferromagnet and
a conventional bulk superconductor (Fig. \ref{fig1}a).
Individually, both systems have an excitation gap,
both in the bulk as well as at the surface.
However, for a special class of antiferromagnetic insulators, as we will discuss below,
protected gapless Andreev bound states emerge at the interface between the two 3D systems. 
These states are mathematically similar to the
Jackiw-Rebbi soliton,\cite{PhysRevD.13.3398}
so that interfacial zero modes exist independently
on how the respective magnitudes and spatial profiles between
the two electronic orders are.
Furthermore, once
intrinsic spin-orbit coupling is introduced,  the interface states open a gap,
giving rise to a topological superconducting state (Fig. \ref{fig1}b).
Therefore, this mechanism shows that
antiferromagnetic insulators, commonly overlooked, are potential candidates to engineer topological superconductors.

The key ingredient for our proposal is the existence of Dirac
lines\cite{PhysRevLett.115.036806,PhysRevLett.95.016405,PhysRevB.84.060504,PhysRevLett.89.077002,PhysRevX.5.011029,hyart2017two,PhysRevLett.116.127202,PhysRevB.93.020506,PhysRevB.97.094508},
lines of points in the Brillouin zone
where the low energy model is a Dirac equation, in
the non-magnetic state of the antiferromagnet.
There is no specific requirement for the superconductor, apart from having a conventional s-wave Cooper pairing.
For the sake of concreteness, we start by introducing a minimal model 
that exemplifies such a phenomenology. For this purpose, we take an antiferromagnetic diamond lattice with lattice constant $a$,
which can be viewed as a three dimensional analog of the antiferromagnetic
honeycomb lattice.\cite{PhysRevLett.114.056403}
Such a structure would be the minimal model for an antiferromagnetic spinel XY$_2$Z$_4$, with the magnetic
ions sitting in the X sites.
\cite{PhysRevX.6.041055,PhysRevB.95.094404,PhysRevB.96.064413,PhysRevB.73.014413,gao2017spiral}  
In order to describe the antiferromagnet-superconductor heterostructure, we propose a Hamiltonian
consisting of electron hopping $H_{kin}$,  antiferromagnetic ordering $H_{AF}$, superconducting s-wave pairing $H_{SC}$,
and spin-orbit coupling $H_{SOC}$\cite{PhysRevLett.98.106803}: $ \hat H = 
\hat H_{kin} +
\hat H_{AF} +
\hat H_{SC} +
\hat H_{SOC} $ with

\begin{equation}
\begin{array}{ll}
\hat H_{kin} &= \sum_{\langle ij \rangle,s} t_{ij} c^\dagger_{i,s} c_{j,s} -  \sum_{i,s} \mu(z_i) c^\dagger_{i,s} c_{i,s}
\\
\hat H_{AF} &= \sum_{i,s,s'} m(z_i) \tau_z^{i,i}
\sigma_z^{s,s'} c^\dagger_{i,s} c_{i,s'}
\\
\hat H_{SC} &= \sum_{i} \Delta(z_i) [
c_{i,\downarrow} c_{i,\uparrow}+
c^\dagger_{i,\uparrow} c^\dagger_{i,\downarrow}]
\\
\hat H_{SOC} &= \sum_{\langle\langle ij \rangle\rangle} i\Lambda 
\vec \sigma^{s,s'}\cdot(\vec r_{il} \times \vec r_{lj})
c^\dagger_{i,s} c_{j,s'}
\end{array}
\label{fullh}
\end{equation}
The parameters are chosen so that the Hamiltonian
describes an insulating antiferromagnet for $z<0$,
with magnetization perpendicular to the interface,
and a conventional superconductor for $z>0$.
In this way, the electronic spectra of the previous Hamiltonian
has an antiferromagnetic
gap for $z=-\infty$ and a superconducting gap for $z=+\infty$.
We may take $m(z) = m_0(1-\tanh(z/W))/2$ the antiferromagnetic order parameter, 
$\Delta(z) = \Delta_0(1+\tanh(z/W))/2$
the superconducting order parameter, 
$\mu(z)=\mu_0(1+\text{sign}(z))/2$ the chemical potential fixing half-filling on the antiferromagnetic side. 
The parameter $W$ controls the smoothness of the change between the two orders,
which in the limit $W\rightarrow 0$ becomes sharp.
Spin-orbit coupling 
enters as a next-nearest neighbor hopping\cite{PhysRevLett.98.106803} between
sites $i$ and $j$,
and $\vec r_{il}$ ($ \vec r_{lj} $) is the vector between
nearest neighbors $i$ ($j$) and $l$.
We denote $\vec \tau$ and $ \vec \sigma$ the Pauli matrices for the
sublattice (A and B) and the spin, respectively. 
The heterostructure within the $fcc$ lattice is chosen so that the interface (perpendicular to the
$z$-axis) consists only of sites belonging to one of the two sublattices, i.e. a zigzag-like interface. 
Using the standard $fcc$ lattice vectors, $ \vec a_1 , \vec a_2 , \vec a_3 $ we can also define the interface plane
by two of them, say $ \vec a_1 $ and $ \vec a_2 $ such that the $z$-axis is parallel to $ \vec a_1 \times \vec a_2 $.  

The first interesting finding is that, in the absence of spin-orbit coupling ($\Lambda =0 $), 
the spectrum of the combined structure develops gapless quasiparticle excitations at the interface [Fig. \ref{fig1}(c)].
These gapless Andreev modes are protected against different choices of the
interface profile for the antiferromagnetic order,
the superconducting order and the chemical potential. 
Due to their robustness and structure shown below, we refer to these protected
Andreev modes as solitonic states. Switching on spin-orbit coupling
$(\Lambda\neq 0)$ leads to a fully gapped spectrum for the solitonic states
[Fig. \ref{fig1}(d)]. The second remarkable observation is the appearance of
the topological Chern invariant $\mathcal{C}=2$ for the gapped system,
indicating the presence of two propagating Majorana modes at the edges of the interface [Fig. \ref{fig1}(e)]. This chiral 
state relies on the broken time reversal symmetry due to the antiferromagnetic order. 
 
The emergence of this topological insulating state by combining two topologically trivial insulating systems
is the main finding of our manuscript.
This topological superconducting state is robust upon changing parameters [Fig. \ref{fig1}(f)],
raising two questions. First, why the interface between the two topologically trivial gapped materials shows
robust zero energy modes? Second, why including a small spin-orbit coupling 
gives rise to a topological superconducting state?

\begin{figure}[t!]
 \centering
                \includegraphics[width=0.95\columnwidth]{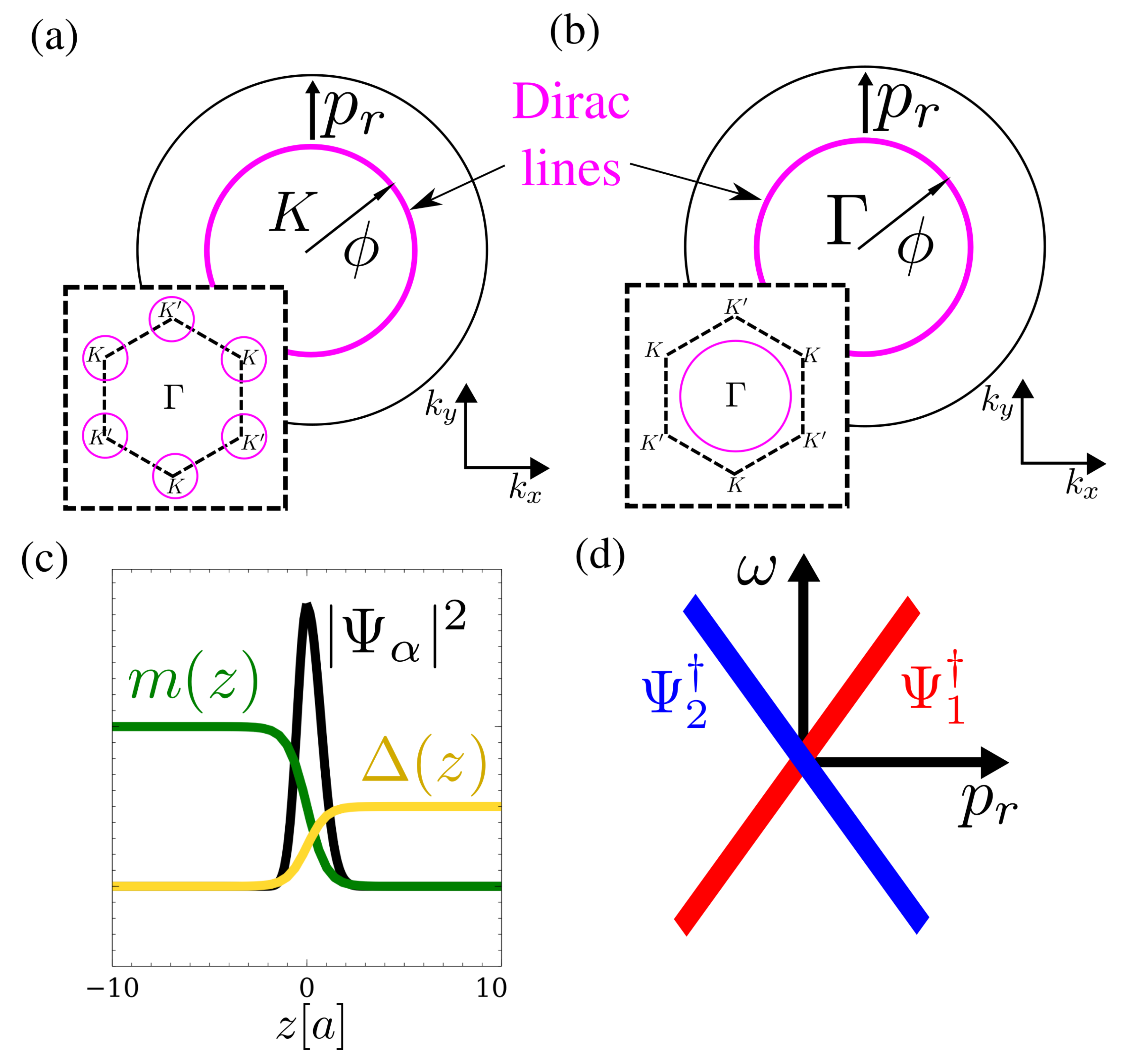}

\caption{
Two structures of Dirac lines projected in the interface plane,
either around the valleys $K $ ($K'$) (a) and around $\Gamma$ (b).
The in-plane momentum is expressed in polar form by $p_r$ and $\phi$,
where $p_r$ denotes the radial distance to the Dirac line and
$\phi$ parametrizes the angle.
For $p_r=0$, each point in the Dirac line will give rise
to the states of Eq.(\ref{soli}),
located at the interface as shown in panel (c). 
The two interface states $\Psi^\dagger_1$
and $\Psi^\dagger_2$ depend on the
radial momentum $p_r$ and angle $\phi$ and
disperse linearly near the Dirac line (d).
}
\label{fig2}
\end{figure}

We first address the origin of the gapless interface states, starting with
the Bloch Hamiltonian for the pristine diamond lattice
$
\hat H_{kin} = \sum_{\vec{k},s}
f(\vec k) c^\dagger_{A,\vec k,s} c_{B,\vec k,s} + 
c.c.
$, where $f(\vec k) = t[1 + 
e^{i\vec k \cdot \vec a_1}+
e^{i\vec k \cdot \vec a_2}]+t'
e^{i\vec k \cdot \vec a_3}
$,
with $\vec a_1, \vec a_2, \vec a_3$ the lattice vectors
of the fcc lattice, 
and $t'=t$ corresponds to the cubic symmetry. The spectrum possesses
lines in $k$-space where the valence and conduction band touch. 
The projected two-dimensional Brillouin zone (BZ) perpendicular to the $z$-axis is hexagonal
with the $ \Gamma $ point (line) in the center and the $ K $ and $ K' $ points (lines) at the boundary. 
Depending on the ratio $ t'/t $, one Dirac line forms around the $ \Gamma $
point or two disconnected Dirac lines form
around $ K $ and $ K' $ points  [Figs. \ref{fig2}(a,b)]. \cite{hyart2017two} 
Focusing on such a Dirac line, we can formulate an effective low-energy Hamiltonian
$\hat H_{D} = 
\sum_{\vec k,s} (p_z - ip_r) c^\dagger_{A,\vec k ,s}  c_{B,\vec k,s} + c.c. .
$
We use that the momentum
$ \vec p $ is tied to the reference frame of the line, such
that $ p_\phi $ is tangential to the line, $ p_r $ perpendicular to the line
and the $z$-axis and $ p_z $ perpendicular to the two other components,
slightly tilted with respect to the $z$-axis. This low-energy model allows us
to study the interface between the superconductor
and antiferromagnet analytically. Using the spatially dependent order
parameters $ \Delta(z) $ and $ m(z) $ 
as introduced above, the effective Hamiltonian takes the form

\begin{equation}
\begin{array}{c@{}l}
\hat H = \sum_{i,j,s}
[\tau_x^{i,j} p_z +
\tau_y^{i,j} p_r
]c^\dagger_{i,\vec k_{\|},s}c_{j,\vec k_{\|}s}  + \\
\sum_{i,s}
m(z)\tau_z^{i,i}\sigma_z^{s,s}c^\dagger_{i,\vec k_{\|},s}c_{i,\vec k_{\|}s}  
+\\
\sum_{i}
\Delta(z) c_{i,-\vec k_{\|},\uparrow} c_{i,\vec k_{\|},\downarrow} + c.c.
\label{interfaceh}
\end{array}
\end{equation}
where $\vec k_\|$
is the conserved Bloch momentum parallel to the interface, 
$i,j$ sum runs over the two sites $A,B$ and $\mu=0$.

\begin{figure}[t!]
 \centering
                \includegraphics[width=0.9\columnwidth]{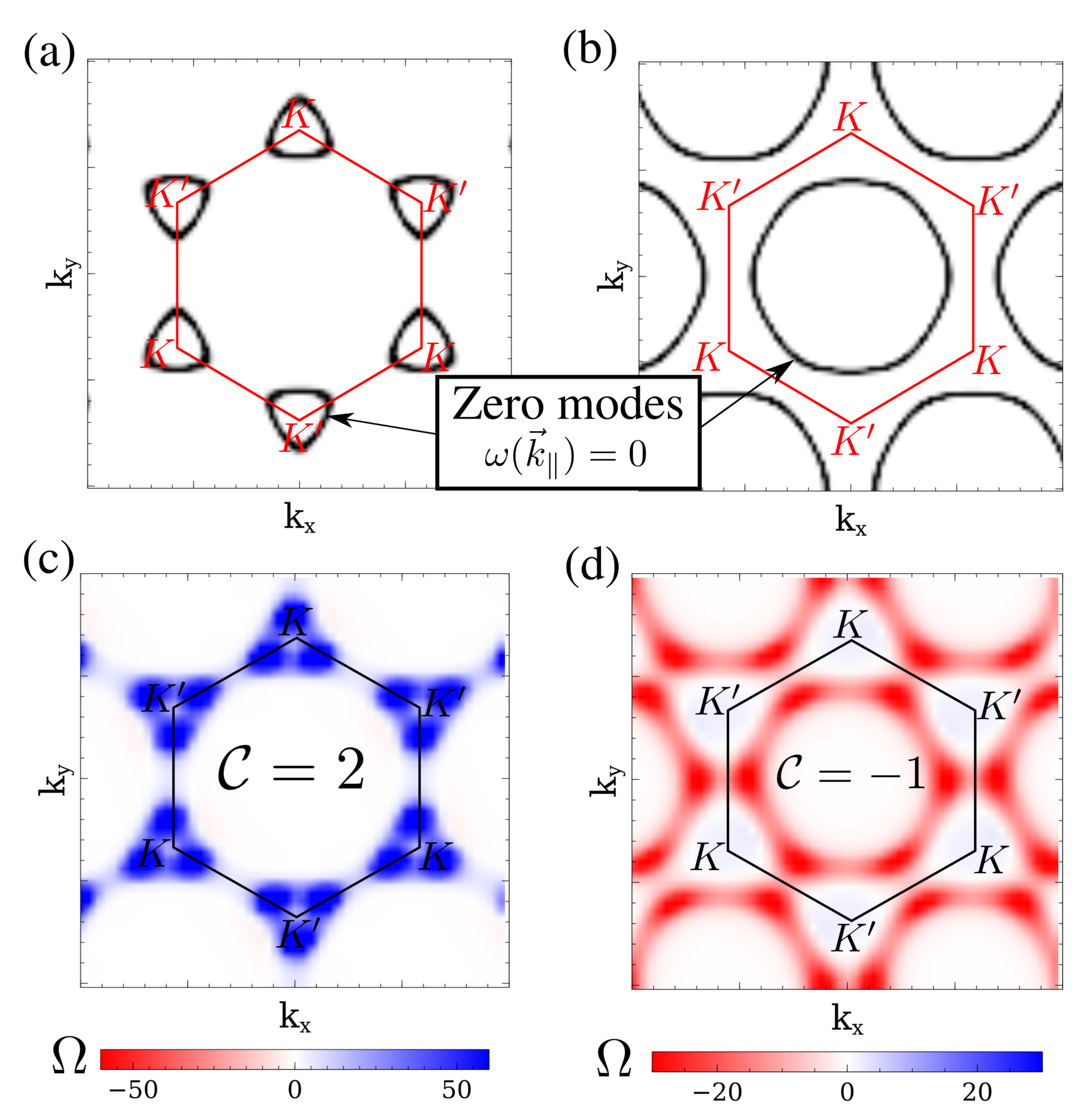}

\caption{(a,b) Zero energy modes
$\omega(\vec k _\|) = 0$
obtained
for the Hamiltonian of Eq. \ref{fullh}
in the absence of spin-orbit.
The transition between the
two states (a,b)
is controlled by the ratio $t'/t$, i.e. one could induce (a) with tensile
($t'/t=0.7$) and (b) with
compressive uniaxial strain ($t'/t=1.5$).  Upon introducing of spin-orbit
coupling, the topological gap opens up,
generating the Berry curvature $\Omega$ localized around the 
former zero modes (c,d).
The two gapped spectra
have a Chern number of $\mathcal{C}=2$ for (c)
and $\mathcal{C}=-1$ for (d).
}
\label{fig3}
\end{figure}

The Hamiltonian (\ref{interfaceh}) defines a system which
is inhomogeneous along the $z$-direction, where for
$z\rightarrow -\infty$ the Hamiltonian is purely antiferromagnetic
and for $z\rightarrow\infty$ purely superconducting.
Remarkably, for $p_r=0$ and a profile fulfilling these asymptotic conditions,
two solitonic zero-energy Andreev modes exist
localized at the interface (Fig. \ref{fig2}c),
with the following ansatz\cite{PhysRevD.13.3398,PhysRevX.5.041042}

\begin{equation}
\begin{array}{l}
\Psi^\dagger_{\alpha,\vec k_{\parallel}}(z) = g(z)
[ c^\dagger_{A,\vec k_{\|},\uparrow} -c_{A,-\vec k_{\|},\downarrow}
\\ \qquad \qquad \qquad \qquad + (-1)^{\alpha} i ( c^\dagger_{B,\vec k_{\|},\uparrow} 
+ c_{B,-\vec k_{\|},\downarrow} )
]
\end{array}
\label{soli}
\end{equation}
where $g(z) = C \exp[\int_0^z[m (z') - \Delta(z')]dz']$,
$C$ as the normalization constant and $ \alpha=1,2 $ as branch index.
Note that although these states
are pinned to zero energy, they are not Majorana modes.
Furthermore, such states will also exist in the
more generic case
$|\Delta(z\rightarrow\infty)|>|m(z\rightarrow\infty)|$
and 
$|m(z\rightarrow-\infty)|>|\Delta(z\rightarrow-\infty)|$.
Away from $p_r=0$, the two solitonic wave functions 
have a finite energy dispersion in the direction of the
radial momentum $p_r$ (Fig. \ref{fig2}d),
yielding the effective Hamiltonian
$\hat H=-\sum_\alpha (-1)^{\alpha} v_rp_r\Psi^\dagger_{\alpha,\vec k_{\|}}
\Psi_{\alpha,\vec k_{\|}}$.
The existence of these states for each point of the Dirac line
implies that the zero mode surface of the heterostructure
reflects the original Dirac lines of the antiferromagnet. Thus, any change
of the Dirac line structure would be reflected in these zero-energy interface
modes, as shown in Figs. \ref{fig3}ab.

\begin{figure}[t!]
 \centering
                \includegraphics[width=\columnwidth]{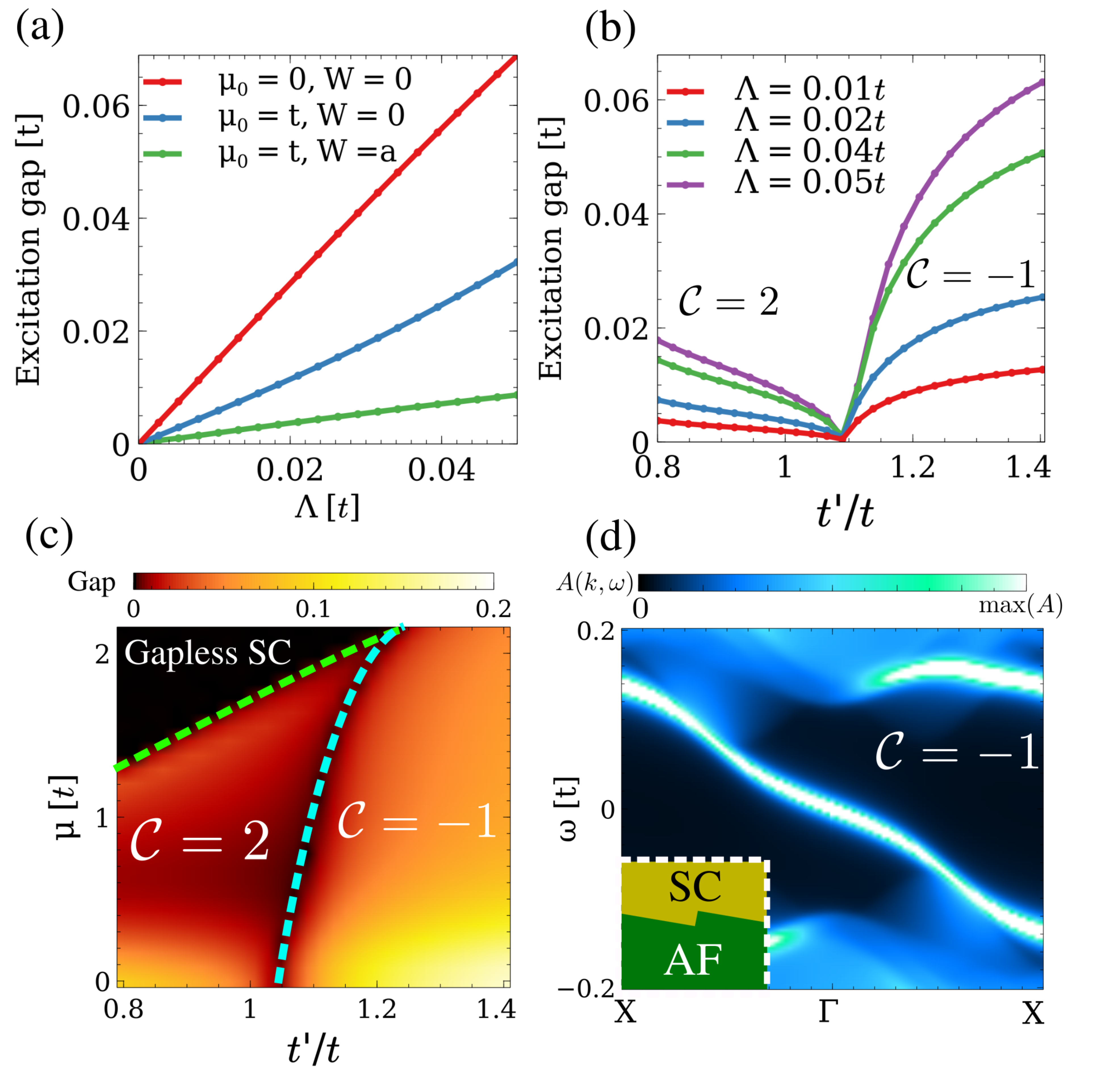}

\caption{
(a) The gap of the topological phase as a function
of spin-orbit coupling calculated for the full model
in Eq. \ref{fullh}, shows a linear scaling in different regimes.
(b) Varying $ t'/t $ in the antiferromagnet drives a topological phase
transition between two topological states, indicated by the gap closing at $ t'/t \approx 1.1 $
 independent of spin-orbit orbit coupling.
(c) Phase diagram for the topological phases
at $\Lambda=0.05t$, $\Delta_0=0.4t$ and $m_0=0.7t$
as a function of the superconducting chemical potential
$\mu$ and $ t'/t $, which shows 
extended gapped regions with $\mathcal{C}=2$, $\mathcal{C}=-1$
and a gapless state for large $\mu$.
The topological superconducting state also arises
in a saw-shaped interface (d), suggesting that 
a perfect interface is not a necessary requirement.
}
\label{fig4}
\end{figure}

In a next step, we introduce the intrinsic spin-orbit coupling, equivalent
to a momentum and sublattice dependent exchange field.
In the vicinity of the
Dirac lines it takes the effective form
$
\hat H_{SOC} \propto
\Lambda
\tau_z^{i,i}
[-\sin(\phi) \sigma_x^{s,s'} + \cos(\phi) \sigma_y^{s,s'}]
c^\dagger_{i,\vec k_{\|},s}
c_{i,\vec k_{\|},s'}
$,
where $ \phi $ denotes the position on the Dirac line  as shown in Fig. \ref{fig2}(a,b).
We note that for the different situations of the Dirac lines the
SOC takes a vortex-like profile, but with opposite vorticities
around $\Gamma$ and $K,K'$.
By projecting the spin-orbit coupling term onto the solitonic basis,
we arrive to the following low-energy Hamiltonian,

\begin{equation}
\mathcal{H}(p_r,\phi) = 
\begin{pmatrix}
\Psi^\dagger_{1,\vec k_{\|}} & 
\Psi^\dagger_{2,\vec k_{\|}}
\end{pmatrix}
\begin{pmatrix}
v_rp_r & -i\lambda e^{i\phi} \\
i\lambda e^{-i\phi} & -v_r p_r \\
\end{pmatrix}
\begin{pmatrix}
\Psi_{1,\vec k_{\|}} \\
\Psi_{2,\vec k_{\|}} \\
\end{pmatrix}
\label{hkink}
\end{equation}
where $\lambda(\Lambda,\mu,W)=\pm|\langle \Psi_1 | H_{SOC} |\Psi_2 \rangle | \propto \Lambda $
generates a gap in the spectra and
finite Berry curvature
where the zero modes were located [Figs.\ref{fig3}(cd)]. 
The gap is linear in the spin-orbit
coupling and depends on the chemical potential and the profile
width $W$ [Fig. \ref{fig4}(a)]. 
This Hamiltonian has the structure of a chiral p-wave superconductor, 
since the spin-orbit coupling $\lambda$ takes the form of a chiral gap
function. 
In this way, the superconducting phase in the interface acquires chirality with a non-vanishing
Chern number, if $\lambda\neq 0$.

The Lifshitz transition
found 
in the paramagnetic phase of the antiferromagnetic side
by varying the hopping ratio $ t'/t $
[Figs\ref{fig2}(a,b)]
has a final consequence for the gapped interface modes:
this
Lifshitz transition introduces a topological transition for the
superconducting phase of the
heterostructure. For $ t' < t $ each of the two Dirac lines contributes
through a single phase winding adding together to a Chern number $\mathcal{C}=
2$ [Fig.\ref{fig3}(c)], while for $ t' > t $ there
is only a single Dirac line winding in opposite
orientation around $ \Gamma $ leading to
$\mathcal{C}= -1$ for the interface superconductor [Fig.\ref{fig3}(d)]. 
Due to corrections to the low energy model,
the topological
phase transition found by exactly solving
the model does not coincide perfectly with the bulk Lifshitz transition,
but happens at $ t' $ slightly higher than $ t $, as visible in Fig.
\ref{fig4}(b), which shows a gap closing at this transition point. 
This specific transition point depends on the chemical potential
$\mu$ as shown in Fig.
\ref{fig4}(c), 
depicting
a phase diagram with two topological phase transitions,
from a gapless superconductor to the topological sector $\mathcal{C}= 2$ and
then  $\mathcal{C}= -1$. Our calculations
demonstrate that the topological
phases are robust, and their existence does not depend on details of the
electronic structure of the superconductor, but is determined by the topology
of the Dirac lines of the magnetic side. Since the symmetric case $ t'/t =1 $
belongs to the sector $\mathcal{C}= 2$, the sector $ \mathcal{C}= -1$ could be
reached through uniaxial strain perpendicular to the interface, increasing $
t'/t $. 

A last important issue, especially for future experimental realizations, is
whether topological phases are sensitive to the quality of the interface. To
test this we now consider a saw-shaped interface, i.e. a tilted interface
orientation yielding a periodicity $(3,1)\times(3,1)$ of the original unit
cell. We observe that even for this ''imperfect'' heterostructure the interface
develops a topological phase with $\mathcal{C}=-1$ for $ t'=t $  [Fig.
\ref{fig4}(d)]. 
The inter-valley scattering induced by the interface supercell shifts the system to the sector $\mathcal{C}=-1$.
Similar results are obtained for other interface orientation, with the
exception of the armchair interface where the two sublattice sites appear in
equal number at the interface. This result demonstrates that the topological
phase can be ascribed to 
the robustness of the parent solitonic states and generically requires an
imbalance between the two sublattice sites. 

Using a minimal model, we have shown how to engineer topological
superconductivity connecting an insulating antiferromagnet with a conventional superconductor.
While we use a single-orbital model, multi-orbital extensions of Eq.
\ref{fullh} could, for example, capture the physics of antiferromagnetic
spinels, such as CoAl$_2$O$_4$, that realizes an insulating antiferromagnetic
diamond lattice,\cite{PhysRevB.95.094404} but it is unclear so far whether this
material generates in the paramagnetic state the necessary Dirac lines.
Finally, it is important to notice that the phenomenology presented here is not
restricted to antiferromagnets on diamond lattices, but will emerge in generic
systems displaying this kind of Dirac lines, which would enlarge the range of
potential candidate
materials.\cite{PhysRevLett.115.036806,PhysRevLett.95.016405,PhysRevB.84.060504,PhysRevLett.89.077002,PhysRevX.5.011029,hyart2017two,PhysRevLett.116.127202}
Such kind of antiferromagnets would constitute an invaluable building block to
engineer two-dimensional topological superconductors
without fine tuning requirements,
robust against imperfections and changes of materials,
providing a new platform to study Majorana physics.

\textbf{Acknowledgements} {We would like to thank 
F. von Oppen, T. Neupert, F. Guinea, A. Ramires
and Y. Maeno
for helpful discussions. 
J.L.L is grateful for financial support from ETH
Fellowship
program. M.S. acknowledges the financial support by the Swiss National Science
Foundation through Division II (No. 163186).
We acknowledge financial support from the 
JSPS Core-to-Core program "Oxide Superspin" international network.

\section*{APPENDIX}

\appendix

\section{Structure of the diamond lattice and the heterostructure}

Here we briefly discuss the spatial structure
of the tight binding scheme employed to model
a Dirac line material. We use a single
orbital tight binding model, whose sites are located in a diamond lattice.
This model can be understood as a three-dimensional extension of 
a honeycomb lattice. The unit cell has two sites, labeled A and B
(Fig. \ref{struct}a). The bulk structure can be understood
as two interpenetrating fcc lattices as shown in Fig. \ref{struct}b.

The heterostructures considered in our manuscript are obtained by growing
the diamond lattice along the (1,1,1) direction
of the $fcc$ lattice shown in Fig. \ref{struct}b.
A sketch of such heterostructure in shown in Fig. \ref{struct}c,
where the purple line mark the interface plane
between the antiferromagnetic and superconducting parts.
The heterostructure can be understood as
stacked buckled honeycomb lattices.
Finally, Fig. \ref{struct}d shows a top view of the interface, showing
a triangular lattice whose Brillouin zone will be hexagonal.

\section{Spectra of isolated AF and SC}
We first briefly discuss the electronic spectra for isolated bulk
superconducting and antiferromagnetic states separately. In the
absence of spin-orbit coupling, both electronic spectra
show a gap, as shown in Fig. \ref{fig1_SM} (a,b), whose magnitude is controlled by $m$
for the antiferromagnet and $\Delta$ for the
superconductor. Including spin-orbit coupling
slightly modifies the spectra but maintains
the gap [Fig. \ref{fig1_SM} (c,d)].
This highlights that spin-orbit coupling does not create a strong change in
the electronic structure for the bulk system.

\begin{figure}[t!]
 \centering
                \includegraphics[width=\columnwidth]{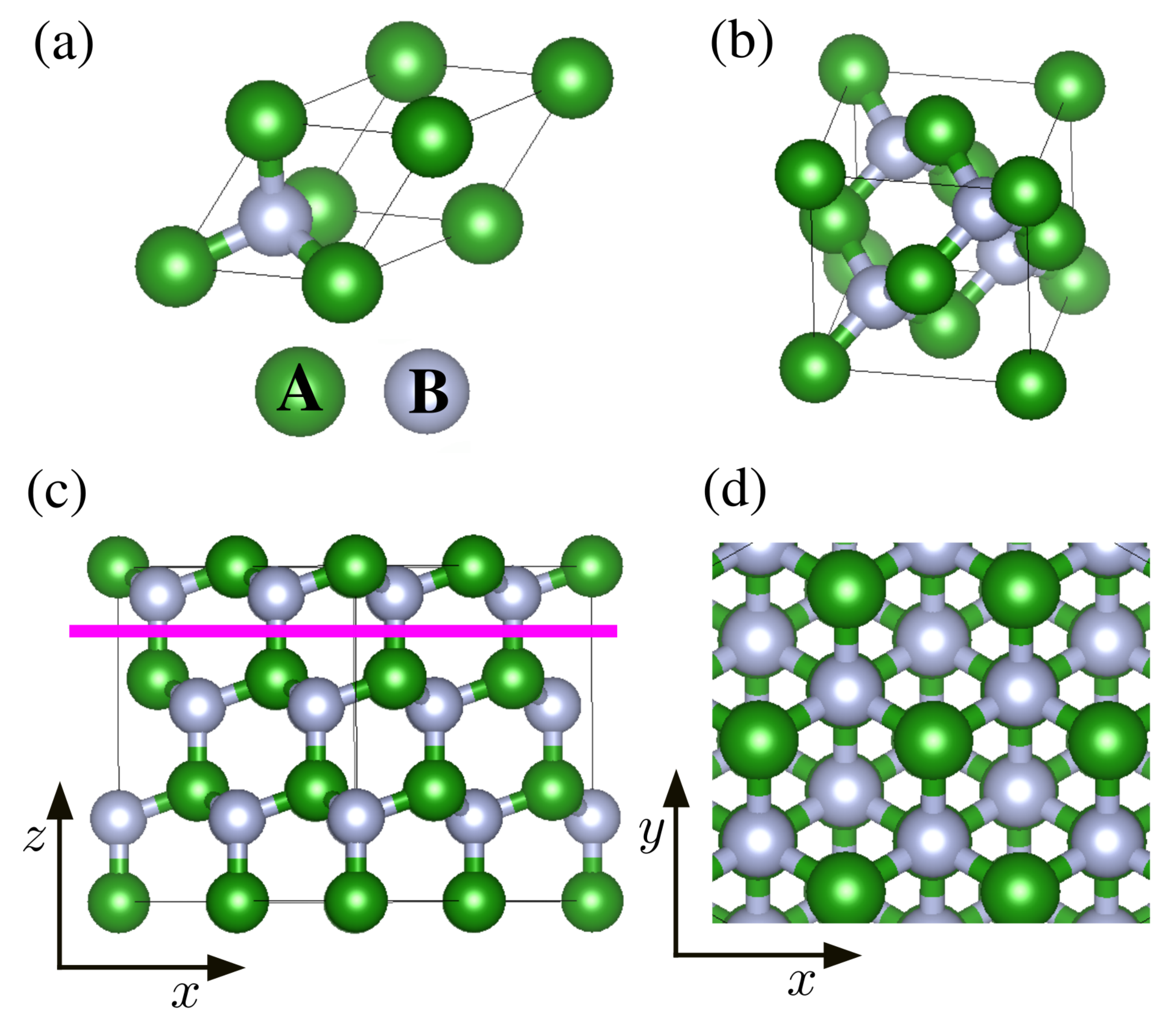}

\caption{ (a) Minimal unit cell of the diamond lattice,
and (b) bulk diamond lattice shown as two interpenetrating
$fcc$ lattices. Panels (c,d)
show a side (c) and top (d) view of the heterostructure
obtained growing along the $fcc$ (1,1,1) direction.
The purple line in (c) marks the interface plane between the
superconductor and the antiferromagnet.
}
\label{struct}
\end{figure}

\section{Spin texture induced by spin-orbit coupling}
We now discuss in more in detail the effect
of spin-orbit coupling,
that enters in the
Hamiltonian as a momentum and sublattice dependent exchange coupling
$\vec m (\vec k)$.
In Fig. \ref{fig2_SM} we show the spin-texture induced by the spin-orbit coupling, projected
onto the Bloch momentum parallel to the interface $\vec k_\|$. 
We observe opposite sign for the spin texture on the two sublattices, naturally connected by
the combination of time reversal and inversion symmetry operation. The exchange field
$\vec m (\vec k)$ induces a vortex structure of the spin textures around the $ K $ and $K'$ point as
well as around the $ \Gamma $ point. This is responsible for the chirality and the structure of
the gapped topological interface states. 
The reciprocal spin texture vanishes in the $M$ point as required by time-reversal symmetry, allowing for
a gap closing in the context of the Lifshitz and the topological phase transition. 

This in-plane spin texture round the $K$, $K'$ and $\Gamma$ points can be captured
by a reduced effective Hamiltonian
of the form
$
\hat H_{SOC} \propto
\tau_z^{i,i}
[-\sin(\phi) \sigma_x^{s,s'} + \cos(\phi) \sigma_y^{s,s'}]
c^\dagger_{i,\vec k_{\|},s}
c_{i,\vec k_{\|},s'}
$
 The winding is visible through the phase factor of the off-diagonal
matrix elements parametrized by the polar angles $\phi^K$ and $ \phi^{\Gamma} $ for a circle around the corresponding centers. Ignoring corrections due to irrelevant details of the band structure, the spin-orbit coupling term can be 
written as 
\begin{equation}
H^{K}_{SO} = \lambda 
\tau_z^{i,i}
[-\sin(\phi^K) \sigma_x^{s,s'} + \cos(\phi^K) \sigma_y^{s,s'}]
c^\dagger_{i,\vec k_{\|},s}
c_{i,\vec k_{\|},s'}
\label{effsoc1}
\end{equation}
in the vicinity of the $ K $ ($K'$) point and
\begin{equation}
H^{\Gamma}_{SO} = -\lambda 
\tau_z^{i,i}
[-\sin(\phi^\Gamma) \sigma_x^{s,s'} + \cos(\phi^\Gamma) \sigma_y^{s,s'}]
c^\dagger_{i,\vec k_{\|},s}
c_{i,\vec k_{\|},s'}
\label{effsoc2}
\end{equation}
around $\Gamma$. Note that the main
the sign change in the effective
exchange field for the two cases, that is visible in the corresponding vorticity 
in Fig. \ref{fig2_SM}.

\begin{figure}[t!]
 \centering
                \includegraphics[width=\columnwidth]{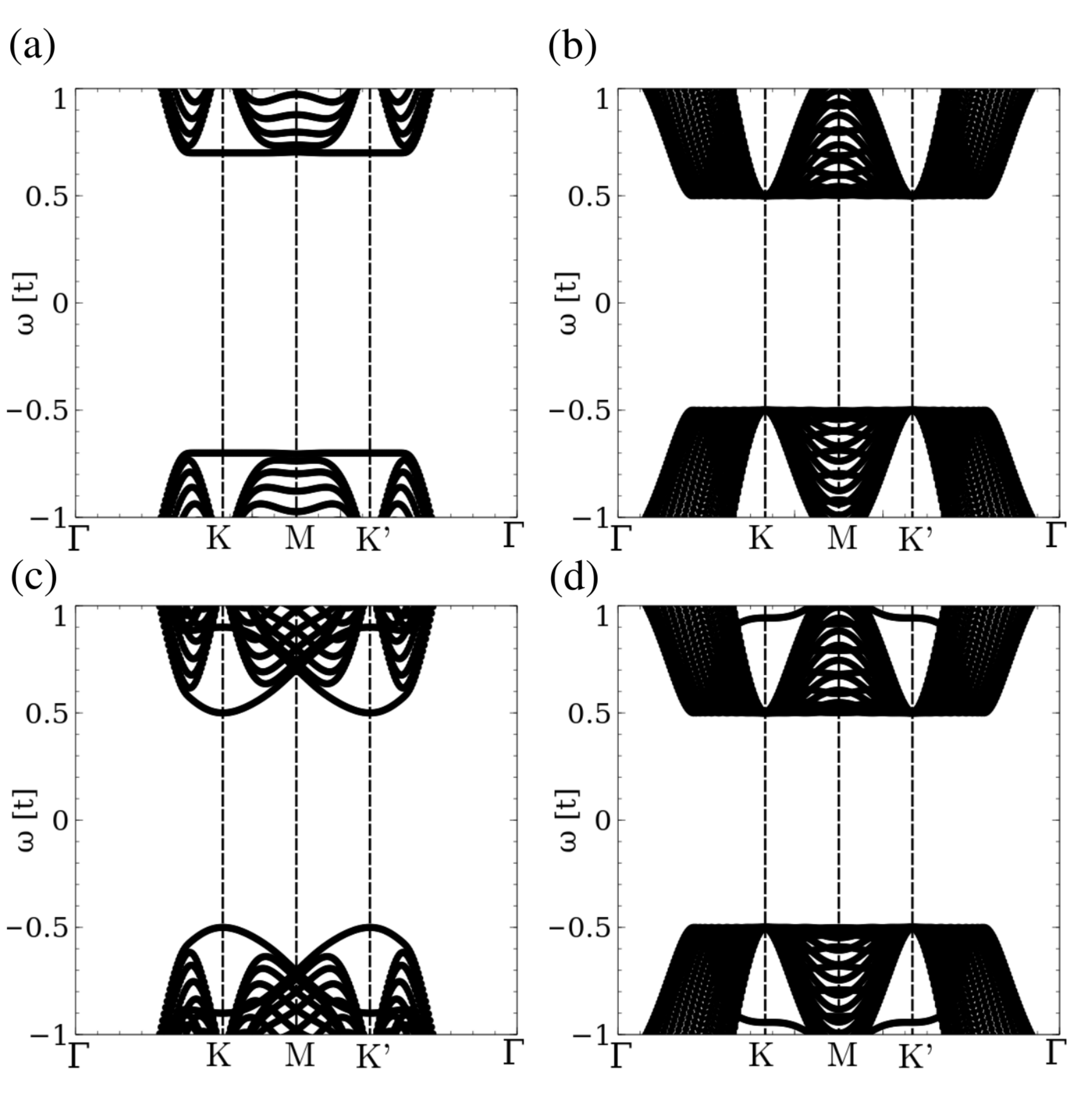}

\caption{ (a) Band structure of the antiferromagnet (a,c) and superconductor
(b,d), without (a,b) and with (c,d) spin-orbit coupling. It
is observed that spin-orbit coupling does not substantially modify
the bulk band-structures.
}
\label{fig1_SM}
\end{figure}

\section{Interface zero-energy states}
We now discuss the analysis of the interface states
based on an effective one-dimensional Dirac equation 
with a spatially dependent antiferromagnetic
ordering and onsite s-wave pairing,
that would arise for the
different $\vec k_\|$ contained in the Dirac line.
This model allows for
a decoupling into two separate sectors, one for spin-up electrons and spin-down holes and another for
spin-down electrons and spin-up holes, with operators $ \varphi_1 $ and $ \varphi_2 $, respectively,

\begin{equation}
\mathcal{H}_1 = 
\varphi^\dagger_1 h_1 \varphi_1
\qquad
\mathcal{H}_2 = 
\varphi^\dagger_2 h_2 \varphi_2
\end{equation}

where 

\begin{equation}
\varphi_1^\dagger = 
\begin{pmatrix}
c^\dagger_{A,{\vec k_{\|}},\uparrow} &
c^\dagger_{B,{\vec k_{\|}},\uparrow} &
c_{A,{-\vec k_{\|}},\downarrow} &
c_{B,{-\vec k_{\|}},\downarrow} 
\end{pmatrix}
\end{equation}

\begin{equation}
\varphi_2^\dagger = 
\begin{pmatrix}
c^\dagger_{A,{\vec k_{\|}},\downarrow} &
c^\dagger_{B,{\vec k_{\|}},\downarrow} &
-c_{A,{-\vec k_{\|}},\uparrow} &
-c_{B,{-\vec k_{\|}},\uparrow} 
\end{pmatrix}
\end{equation}

\begin{equation}
h_1 = 
\begin{pmatrix}
m & p_z & \Delta & 0 \\
p_z & -m & 0 & \Delta \\
\Delta & 0 & m & -p_z \\
0 & \Delta & -p_z & -m \\
\end{pmatrix}
\end{equation}

\begin{equation}
h_2 =
\begin{pmatrix}
-m & p_z & \Delta & 0 \\
p_z & m & 0 & \Delta \\
\Delta & 0 & -m & -p_z \\
0 & \Delta & -p_z & m \\
\end{pmatrix}
\end{equation}
at $ p_r = 0 $. 
The superconducting and antiferromagnetic ordering are
general functions of $z$, so that
$m=m(z)$ and $\Delta=\Delta(z)$. We will impose the
asymptotic conditions $m(-\infty)=m_0$, $m(+\infty)=0$,
$\Delta(-\infty)=0$ and $\Delta(+\infty)=\Delta_0$, with $\Delta_0$
and $m_0$ positive real numbers.
Taking $p_z = -i \partial_z$, 
the following two spinors $u_1,u_2$

\begin{equation}
u_1(z) = 
\frac{1}{2}
\begin{pmatrix}
1 \\
-i\\
-1\\
-i 
\end{pmatrix}
e^{\int_0^z[m (z') - \Delta(z')]dz'}
\end{equation}

\begin{equation}
u_2(z) = 
\frac{1}{2}
\begin{pmatrix}
1\\
i \\
1 \\
-i
\end{pmatrix}
e^{\int_0^z[m (z') - \Delta(z')]dz'}
\label{solwaves}
\end{equation}
are eigenvectors fulfilling $h_1 u_1 = 0$
and $h_2 u_2 = 0$ as expected for a zero-energy state. 
It is important to note that such zero-energy solutions exist for
any profile $m(z)$ and $\Delta(z)$ provided the asymptotic
conditions are fulfilled, and can be understood as solitonic
solutions between a Dirac antiferromagnet and a Dirac superconductor.
In terms of field operators, the zero-energy solutions take the  simple product form

\begin{figure}[t!]
 \centering
                \includegraphics[width=\columnwidth]{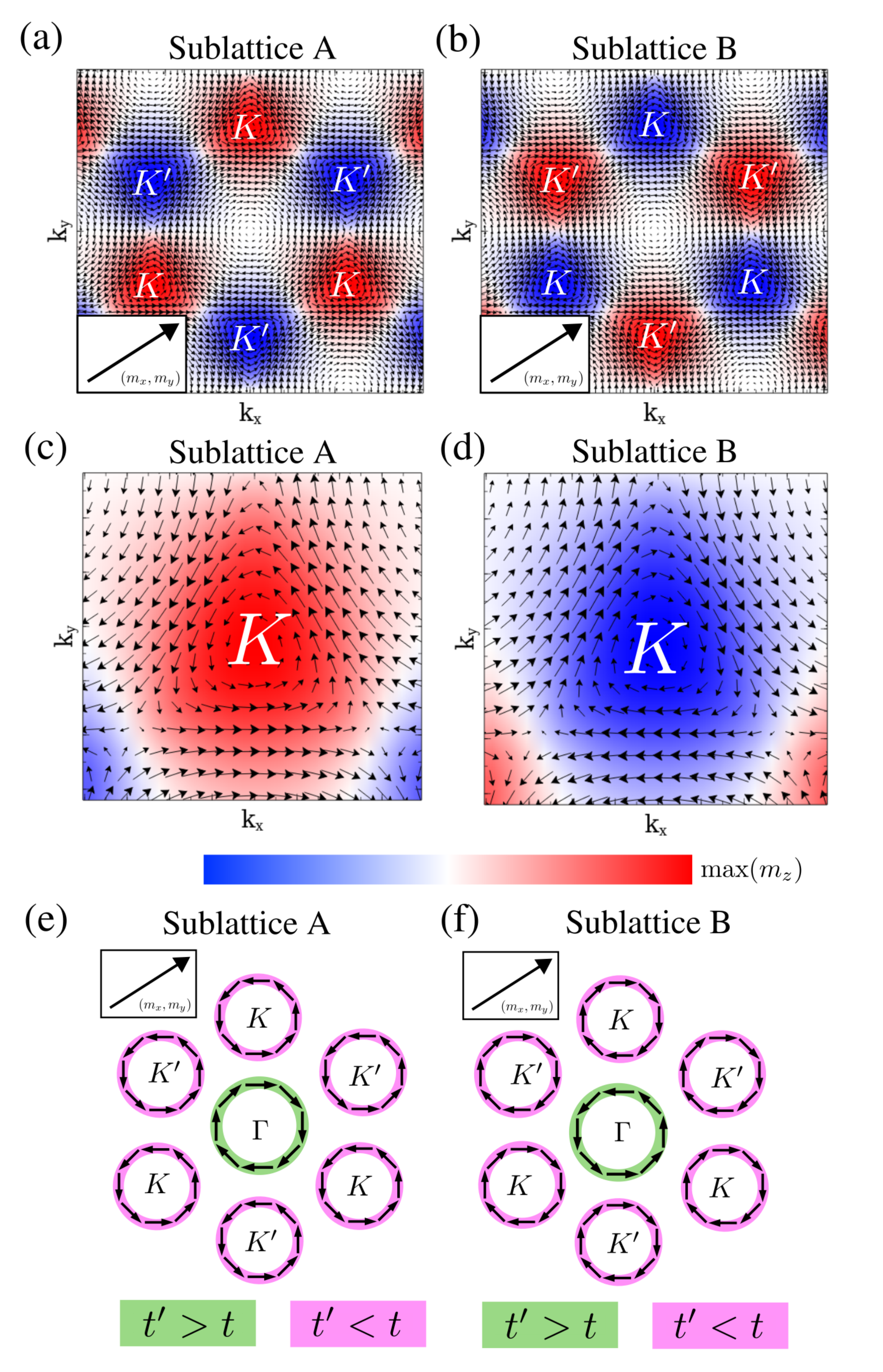}

\caption{ Expectation value of the spin in reciprocal space
for the spin-orbit coupling, for a finite
slab grown along $\vec a_3$. Arrows denote
the spin components
projected in the plane defined
by $a_1$ and $a_2$, whereas the color denotes
the perpendicular component.
The expectation values are 
projected onto the A sublattice (a,c,e)
and B sublattice (b,d,f). It is clearly observed that spin texture changes
between valleys as required by time reversal symmetry, and
between sub-lattices as required by inversion symmetry.
Panels (e,f) show a minimal sketch of the
vortex-like exchange field around $K,K'$ and $\Gamma$.
}
\label{fig2_SM}
\end{figure}

\begin{equation}
\begin{array}{c@{}l}
\Psi^\dagger_1(z) = g(z)
[
c^\dagger_{A,\vec k_{\|},\uparrow} 
-ic^\dagger_{B,\vec k_{\|},\uparrow} 
-c_{A,-\vec k_{\|},\downarrow} 
-ic_{B,-\vec k_{\|},\downarrow} 
]
\\
\Psi^\dagger_2(z) = g(z)
[
c^\dagger_{A,\vec k_{\|},\downarrow} +
ic^\dagger_{B,\vec k_{\|},\downarrow} 
-c_{A,-\vec k_{\|},\uparrow} 
+ic_{B,-\vec k_{\|},\uparrow}
] 
\end{array}
\end{equation}
with $g(z) = C e^{\int_0^z[m (z') - \Delta(z')]dz'}$
and $C$ a normalization constant as stated in the main text.
We finally note that a zero energy solution
generically
exists provided $\Delta(\infty)>m(\infty)$
and $\Delta(-\infty)<m(-\infty)$.
Therefore, our proposal will also hold in the presence of both
superconductivity and antiferromagnetism in the heterostructure,
as long as for $z\rightarrow -\infty$ the antiferromagnetic gap
is bigger than the superconducting gap and
for $z\rightarrow -\infty$
the superconducting gap
is bigger than the antiferromagnetic gap.

This calculation assumes $\mu=0$
in the superconducting part. For a generic situation
with $\mu\neq0$, our numerical calculations
show that the zero energy states still appear, but
at $\vec k_\|$ slightly shifted from the Dirac lines.

\section{Gap opening in interfacial states}
Next we consider possible terms in the Hamiltonian that could
open a gap in the interfacial edge states. 
For that, we will consider several one-body perturbations, and we will
project them onto the solitonic subspace. 
We take a basis that accounts for the two interfacial states as they will not be independent anymore,
\begin{equation}
\varphi = 
\begin{pmatrix}
c_{A,\vec k_{\|},\uparrow} \\
c_{B,\vec k_{\|},\uparrow} \\
c^\dagger_{A,-\vec k_{\|},\downarrow} \\
c^\dagger_{B,-\vec k_{\|},\downarrow} \\
c_{A,\vec k_{\|},\downarrow} \\
c_{B,\vec k_{\|},\downarrow} \\
-c^\dagger_{A,-\vec k_{\|},\uparrow} \\
-c^\dagger_{B,-\vec k_{\|},\uparrow} \\
\end{pmatrix}
\end{equation}

In this basis, the two interface states are now represented as
$\Psi^\dagger_\alpha = \varphi^\dagger\psi_\alpha$ with

\begin{equation}
\psi_1 =
\frac{1}{2}
\begin{pmatrix}
1 \\
-i \\
-1 \\
-i \\
0 \\
0 \\
0 \\
0 \\
\end{pmatrix}
\quad
\psi_2 =
\frac{1}{2}
\begin{pmatrix}
0 \\
0 \\
0 \\
0 \\
1 \\
i \\
1 \\
-i \\
\end{pmatrix}
\end{equation}

such that the projection operator $P$ in the manifold
$(\Psi_1^\dagger ,\Psi_2^\dagger ) = \Psi^\dagger
= \varphi^\dagger P $  takes the form

\begin{equation}
P = 
\frac{1}{2}
\begin{pmatrix}
1 & 0\\
-i & 0\\
-1 & 0\\
-i & 0\\
0 & 1\\
0 & i\\
0 & 1\\
0 & -i\\
\end{pmatrix}
\end{equation}

Given a certain perturbation 
of the form $\mathcal{V} = \varphi^\dagger V \varphi$ in the original basis, its
representation in the solitonic basis,
$\mathcal{W}=\Psi^\dagger W\Psi$, is obtained as
$W=P^\dagger V P$. 
With the previous representation, projecting the different operators
simply consists of multiplying the relevant matrix elements.
The first perturbation that we will consider is a sublattice
independent exchange field, which takes the form

\begin{equation}
\mathcal{U}(\vec k_{\|}) = 
\sum_{i=A,B}
f(\vec k_{\|})
c^\dagger_{i,\vec k_{\|},\uparrow}
c_{i,\vec k_{\|},\downarrow} + c.c
\end{equation}
with
$
f(\vec k_{\|})
$
a generic function. Projecting this in the matrix 
representation yields immediately zero due to the
sublattice structure of $\Psi$. Hence no 
sublattice independent local exchange can break the degeneracy
of the solitonic states, at least to first order. In particular, this 
implies that an external in-plane magnetic field and  a sublattice-independent Rashba spin-orbit coupling 
will not open up a gap.

The next perturbation that we will consider is a sublattice dependent exchange
field, even in momentum, 

\begin{equation}
\mathcal{V}(\vec k_{\|}) =  
g(\vec k_{\|})[
c^\dagger_{A,\vec k_{\|},\uparrow}
c_{A,\vec k_{\|},\downarrow} 
-c^\dagger_{B,\vec k_{\|},\uparrow}
c_{B,\vec k_{\|},\downarrow}]
+ c.c
\end{equation}
with
$
g(\vec k_{\|}) =
g(-\vec k_{\|}) 
$. Once more the projection on the solitonic states is zero, but
now the reason lies in the relative phases between electron and
hole sectors. This perturbation would arise from rotating the axis of the
staggered moment of the antiferromagnet. The degeneracy results from the
spin rotational symmetry of the energy spectra.

Finally, we consider a sublattice dependent exchange field that is odd in momentum,

\begin{equation}
\mathcal{W}(\vec k_{\|}) = 
h(\vec k_{\|})[
c^\dagger_{A,\vec k_{\|},\uparrow}
c_{A,\vec k_{\|},\downarrow} 
-c^\dagger_{B,\vec k_{\|},\uparrow}
c_{B,\vec k_{\|},\downarrow}]
+ c.c
\end{equation}
with $ h(\vec k_{\|}) = - h(-\vec k_{\|}) $. In this case the projection
does not vanish as it fits to the relative phase structure of electrons and holes. 
Such an odd-momentum exchange term arises from spin-orbit coupling, which includes
the sign change between the sublattices due to inversion symmetry. 
To summarize, a perturbation $\mathcal{W}(\vec k_{\|})$ opening a gap in the
solitonic states must fulfill the following conditions

\begin{equation}
\begin{array}{c@{}l}
\hat \Theta : \quad \mathcal{W}(\vec k_{\|}) \rightarrow -\mathcal{W}(-\vec k_{\|}) \\
\hat \Theta \hat P : \quad \mathcal{W}(\vec k_{\|}) \rightarrow \mathcal{W}(\vec k_{\|}) \\
\end{array}
\end{equation}
where $\hat \Theta$ and and $\hat P$ are time reversal and inversion symmetry
operators.

\section{Calculation of the Chern number}
We start with the effective Hamiltonian for the low-energy
Andreev modes around the Dirac lines $\mathcal{H}(p_r,\phi) =
\Psi^\dagger h \Psi$ with

\begin{equation}
h = 
\begin{pmatrix}
v_rp_r & -i\lambda e^{i\phi} \\
i\lambda e^{-i\phi} & -v_r p_r \\
\end{pmatrix}
\end{equation}

where $p_r \in (-\infty,\infty)$ and $\phi \in (0,2\pi)$. By performing a
change of variable $\cos{\theta} = \frac{v_r p_r}{\sqrt{\lambda^2+v_r^2 p_r^2}}$the Hamiltonian can be rewritten as
\begin{equation}
\tilde h = f(\theta)
\begin{pmatrix}
\cos{\theta} & -i\sin{\theta}e^{i\phi} \\
i\sin{\theta} e^{-i\phi} & -\cos{\theta} \\
\end{pmatrix}
\end{equation}
with $f(\theta)>0$ and $\theta \in (0,\pi)$.
This Hamiltonian describes a skyrmion in reciprocal space
with the variables $\theta$ and $\phi$.
The Berry curvature associated with the low-energy state
of this Hamiltonian corresponds to a magnetic monopole in reciprocal space.
Thus, the calculation of the Chern number simply yields the charge of the monopole $\pm 1$.
This leads to a Chern number $\mathcal{C}_{K,K'} =1$ for the $ K $ and $ K' $ point which add up to $\mathcal{C}=2$. 
For the $ \Gamma $-point the skyrmion is reversed leading to the Chern number $\mathcal{C}=-1$. The topological phase transition connects the phase with $\mathcal{C}=2$ and $\mathcal{C}=-1$.

\begin{figure}[t!]
 \centering
                \includegraphics[width=\columnwidth]{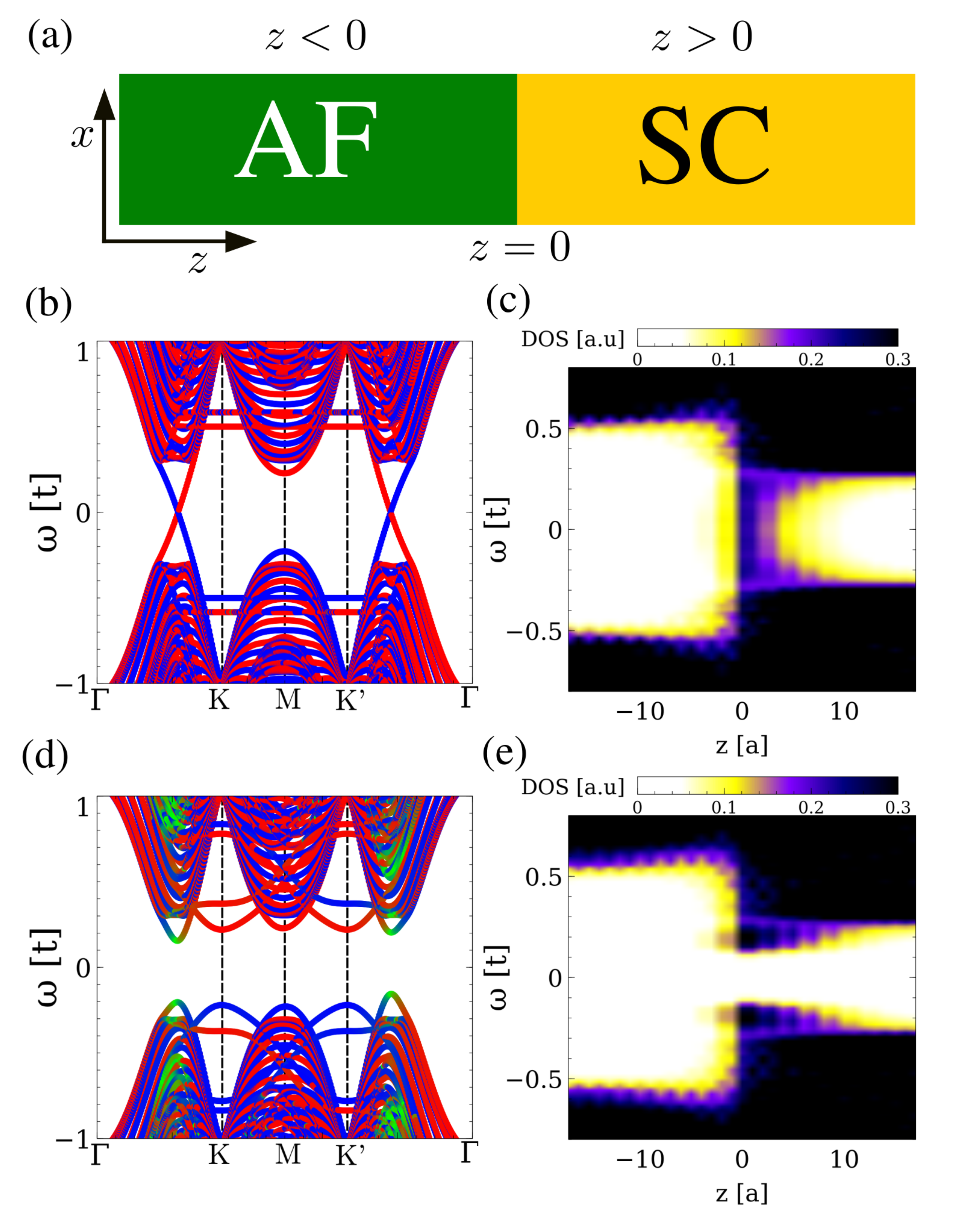}

\caption{
(a) Sketch of the antiferromagnet/superconducting heterostructure
along the perpendicular z direction, taking the interface at $z=0$.
(b) Band structure in the gapless regime $\Lambda=0$
and local density of states along the $z$ direction, showing
a gap closing at the interface where the gapless modes
are located (c).
Upon introduction of spin-orbit coupling, the interface
modes open up a gap (d), giving rise to a gapped spectra
in any point of the heterostructure as shown in (e).
The parameters used are $m_0=0.5t$, $\Delta_0=0.3t$, $W=0$
in (b,c,d,e), $\Lambda=0$ in (b,c) and $\Lambda=0.07t$ in (d,e).
}
\label{dos}
\end{figure}

\section{Spatial dependence of the gap}

In this section we show how the topological
gap evolves as one moves in the
heterostructure from
superconducting to the antiferromagnetic part
in the $z$ direction.
In the topological
state, the gap remains open as one goes away from the
interface in the z-direction.
In the direction of the antiferromagnet the gap converges to $m_0$,
whereas in the direction of the superconductor
it converges to $\Delta_0$.
To rationalize this,
it is illustrative
to compute the density of states (DOS)
the heterostructure as shown in Fig. \ref{dos}a,
where show two different situations: a gapless state
which arises for zero spin-orbit coupling $\Lambda=0$
(Figs. \ref{dos}bc), and gapped situation that
arises when taking $\Lambda\neq 0$ (Figs. \ref{dos}de).
The computation of the density of states
can be performed by means of the Green function of the
heterostructure as
$\text{DOS}(z,\omega)
\propto \text{Im} ( \int_{BZ} G(\omega,\vec k,z) d^2 \vec k )$,
with $G(\omega,\vec k,z)$ the Green function of
the heterostructure, $\omega$ the energy, $\vec k$ the in-plane Bloch
momenta and $z$ the vertical coordinate in the heterostructure.

Both in the absence
($\Lambda=0$)
and
presence
($\Lambda\neq 0$)
of spin-orbit coupling, it is observed that in the
antiferromagnetic region $z<0$, the local gap converges to the antiferromagnetic
gap $m_0$, whereas in the superconducting region $z>0$
it converges to the superconducting gap $\Delta_0$
(Figs. \ref{dos}ce). In the absence of
spin-orbit coupling
(Figs.  \ref{dos}bc), the density of states at the interface $z=0$
becomes gapless, signaling the existence of the gapless modes in that region.
In comparison, for $\Lambda\neq0$ (Figs.  \ref{dos}de)
the density of states remains gapped at the interface.
Therefore, in the case of a topological gap (Figs.  \ref{dos}bc),
the system remains fully gapped for every point of the heterostructure.

\section{Strain driven topological phase transition}

\begin{figure}[t!]
 \centering
                \includegraphics[width=\columnwidth]{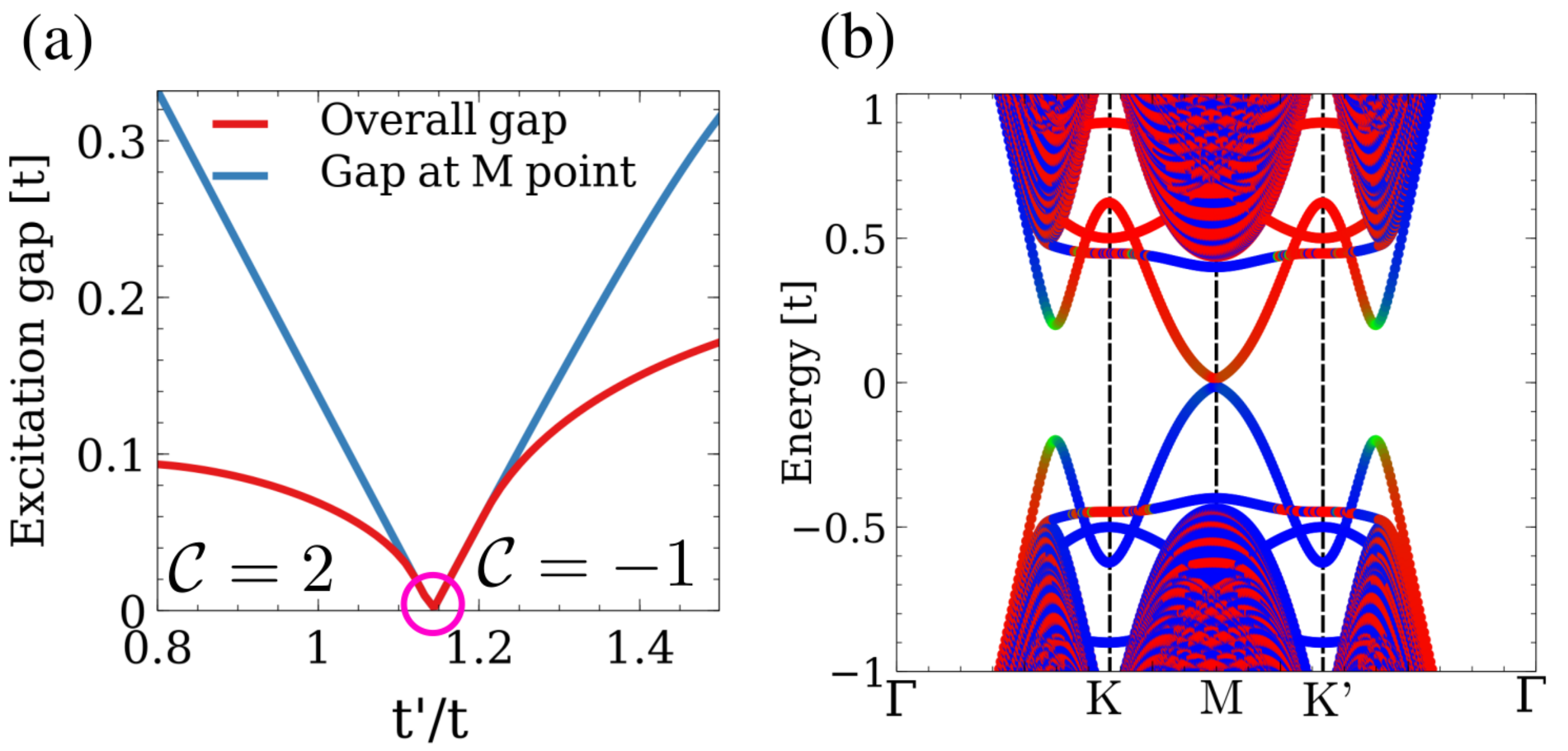}

\caption{ (a)
Evolution of the gap of the heterostructure
with the uniaxial strain $t'/t$, showing
both the overall gap (red) as well as the gap just at the M point (blue).
It is observed that the closing of the overall gap happens through a closing
at the M point.
Panel (b) shows the band structure of the heterostructure
at the gap closing point marked with a purple circle in (a).
It is observed that the interface modes remain gapless at M.
The parameters taken are $m_0 = 0.5t$, $\Delta_0=0.4t$,
$\Lambda = 0.05 t$, $\mu=0$ and $W=0$.
}
\label{gapM}
\end{figure}

Here we briefly discuss
details on the gap closing
for the topological phase
transition with strain between
$\mathcal{C} = 2$
and
$\mathcal{C} = -1$.
The gap closing occurs at the M points,
where the spin-orbit coupling vanishes for symmetry reasons.
To illustrate this, we show
in Fig. \ref{gapM}a the evolution of the
gap with uniaxial strain, both in the M point and in the full Brillouin
zone.
In this way, when the interfacial zero modes are located at the
M point,
that happens at the critical strain
marker with a purple circle
in Fig. \ref{gapM}a,
the system remains gapless even in the presence of spin-orbit
coupling, as shown in
Fig. \ref{gapM}b.
For strains in which there is a finite gap, the gap is
generically not located
at the M point but at some arbitrary point in the
Brillouin zone, close to the location of the Dirac lines.

\section{Possible candidate materials}

The main limitation of our model is that
we cannot unequivocally assess if a specific
complex material
would be suitable for our proposal.
An analogous analysis to the one presented in our manuscript
could be performed by
combining density functional theory and
Wannierization,\cite{RevModPhys.84.1419}
which allows obtaining a multiorbital Hamiltonian
from first principles.
\cite{PhysRevLett.107.186806,PhysRevLett.110.116802,PhysRevMaterials.2.014202}
In that way,
it would be possible to asses from first principles if a certain material
would be suitable for the mechanism presented.

As mentioned in the manuscript, a possible
candidate for our proposal is CoAl$_2O_4$, that is experimentally
known to realize an antiferromagnetic diamond lattice,
giving rise to a multiorbital version of the tight binding model
of our manuscript.
Interestingly, spinels compounds
have been proposed\cite{PhysRevLett.112.036403,PhysRevLett.108.146601,PhysRevLett.107.186806}
to show Dirac and Weyl like crossings,
in particular
CaOs$_2$O$_4$  \cite{PhysRevLett.108.146601},
SrOs$_2$O$_4$ \cite{PhysRevLett.108.146601}
and 
HgCr$_2$Se$_4$\cite{PhysRevLett.107.186806}.
Assessing whether if CoAl$_2O_4$ has Dirac lines in the
paramagnetic state requires first principles density functional
theory calculations, which is beyond the scope of our study.
Nevertheless, given that similar compounds are known to show Dirac-like
physics, it is likely that CoAl$_2O_4$ could realize the required electronic
structure for our proposal.

Assuming that CoAl$_2$O$_4$ hosts the necessary gapped Dirac lines,
it is still necessary to assess the value of the topological gap,
controlled by spin-orbit coupling of the antiferromagnetic and superconductor.
In the following
we will take as the antiferromagnet CoAl$_2$O$_4$
and as superconductor the spinel compound
LiTi$_2$O$_4$\cite{PhysRevB.70.054519,jin2015anomalous},
that has a superconducting gap $\Delta_0 = 1.9$ meV.
In a low energy Hamiltonian, the effective
spin-orbit coupling can be reduced by an order
of magnitude with respect to the atomic value
due to its interplay with crystal field effects.\cite{xiao2011interface}
Taking into account that the atomic spin-orbit coupling in 3d
transition metals is on the order of 20 meV,\cite{PhysRevLett.115.237202}
we would have $\Lambda = 2$ meV
for the effective low energy Hamiltonian.
Therefore, according to the previous discussion for
a CoAl$_2$O$_4$/LiTi$_2$O$_4$
heterostructure,
we may expect a topological gap on the order of 0.4 meV,
which can be observed experimentally
and is on the same order of magnitude
of state-of-the-art
experiments\cite{zhang2018quantized,zhang2017ballistic}.

\bibliographystyle{apsrev4-1}
\bibliography{biblio}{}

\end{document}